\documentclass[letterpaper,twocolumn,10pt]{article}
\usepackage{usenix,epsfig,endnotes}
\usepackage{nohyperref}
\usepackage[hyphens]{url}
\usepackage[linesnumbered]{algorithm2e}
\usepackage{graphicx}
\usepackage{subfigure}
\usepackage{alltt}
\usepackage{array}
\usepackage{color}

\usepackage{authblk}
\newcolumntype{L}[1]{>{\raggedright\let\newline\\\arraybackslash\hspace{0pt}}m{#1}}
\newcolumntype{C}[1]{>{\centering\let\newline\\\arraybackslash\hspace{0pt}}m{#1}}
\newcolumntype{R}[1]{>{\raggedleft\let\newline\\\arraybackslash\hspace{0pt}}m{#1}}
\newcommand{\comment}[1]{}

\begin{document}
\date{}
\title{\Large \bf Website-Targeted False Content Injection by Network Operators}

\author[1,3]{Gabi Nakibly}
\author[2]{Jaime Schcolnik}
\author[3]{Yossi Rubin}
\affil[1]{\small Computer Science Department, Technion, Haifa, Israel}
\affil[2]{\small Computer Science Department, Interdisciplinary Center, Herzliya, Israel}
\affil[3]{\small Rafael -- Advanced Defense Systems, Haifa, Israel}


\maketitle

\subsection*{Abstract}

It is known that some network operators inject false content into users' network traffic. Yet all previous works that investigate this practice focus on edge ISPs (Internet Service Providers), namely, those that provide Internet access to end users. Edge ISPs that inject false content affect their customers only. However, in this work we show that not only edge ISPs may inject false content, but also  core network operators. These operators can potentially alter the traffic of \emph{all} Internet users who visit predetermined websites. We expose this practice by inspecting a large amount of traffic originating from several networks. Our study is based on the observation that the forged traffic is injected in an out-of-band manner: the network operators do not update the network packets in-path, but rather send the forged packets \emph{without} dropping the legitimate ones. This creates a race between the forged and the legitimate packets as they arrive to the end user. This race can be identified and analyzed. Our analysis shows that the main purpose of content injection is to increase the network operators' revenue by inserting advertisements to websites. Nonetheless, surprisingly, we have also observed numerous cases of injected malicious content. We publish representative samples of the injections to facilitate continued analysis of this practice by the security community.

\section{Introduction}

Over the last few years there have been numerous reports of ISPs that alter or proxy their customers' traffic, including, for example, CMA Communications in 2013 \cite{CMA}, Comcast in 2012 \cite{comcast}, Mediacom in 2011 \cite{mediacom},  WOW! in 2008 \cite{Topolski2008}, and Rogers in 2007 \cite{rogers}. Moreover, several extensive studies have brought the details of this practice to light \cite{kreibich2010netalyzr, weaver2014here, reis2008detecting, Zimmerman2015}. The main motivations of ISPs to alter traffic are to facilitate caching, inject advertisements into DNS and HTTP error messages, and compress or transcode content. 

\begin{figure*}
\begin{center}
	\subfigure[Depiction of monitored traffic in the server-centric approach (of past works). One server with specific content serves many clients in many edge networks.]{
		\label{fig:servercentric} 
		\includegraphics[width=0.36\textwidth]{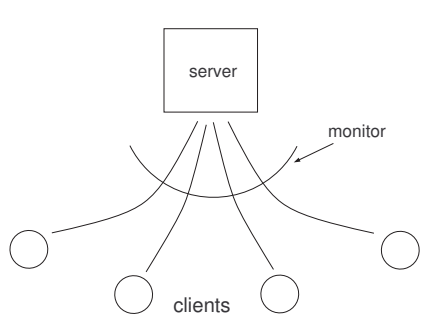}}
		\hspace{1cm}
	\subfigure[Depiction of monitored traffic in the client-centric approach (of the current work). Many servers with varied content serve many clients in a few edge networks.]{
		\label{fig:clientcentric} 
		\includegraphics[width=0.4\textwidth]{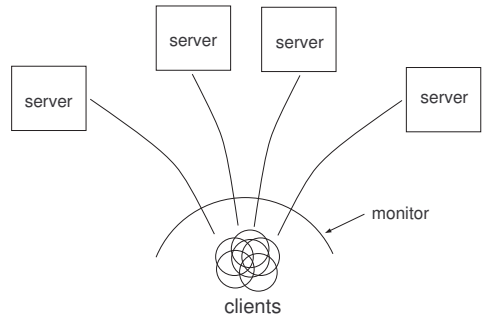}}
	\caption{Server-centric approach versus client-centric approach to monitoring traffic. The lines between clients and servers illustrate the monitored traffic.}
	\label{fig:appraoches} 
\end{center}
\vspace{-10pt}
\end{figure*}

All of these reports and studies found that  these traffic alterations were carried out exclusively by \emph{edge ISPs}, namely, retail ISPs that sell Internet access directly to end customers, and are their ``first hop" to the Internet. This finding stems from the server-centric approach the above studies have taken. In this approach, one or a handful of servers are deployed to deliver specific content to users, after which a large number of clients are solicited to fetch that content from the servers. Finally, an agent on the clients -- usually a JavaScript delivered by the server itself -- looks for deviations between the content delivered by the server and that displayed to the user. Figure~\ref{fig:servercentric} illustrates the traffic monitored in this server-centric approach.
 
Such an approach can be used to inspect the traffic of many clients from diverse geographies who are served by different edge ISPs. The main disadvantage of this approach is that the content fetched by the clients is very specific. All clients fetch the same content from the same web servers. This allows only the detection of network entities that aim to modify all of the Internet traffic\footnote{In some cases these network entities modify all internet traffic originating from very popular websites such as \url{google.com}, \url{apple.com}, and \url{bing.com} or all Internet traffic originating from \url{.com}.} of a predetermined set of users and are generally oblivious to the actual content delivered to the user. Such entities indeed tend to be edge ISPs that target only the traffic of their customers.

In this work we show that the above approach misses a substantial portion of the on-path entities that modify traffic on the Internet. 
Using extensive observations over a period of several weeks, we analyzed petabits of Internet traffic carrying varied content delivered by servers having over 1.5 million distinct IP addresses. We newly reveal several network operators that modify traffic not limited to a specific set of users. Such network operators alter Internet traffic on the basis of its content, primarily by the website a user visits. The traffic of \emph{every} Internet user that traverses these network operators is susceptible to alteration. This is in contrast to the case of edge ISPs that alter the traffic of their customers only. Although a primary focus of these network operators is to inject advertisements into web pages, we also identified injections of malicious content. 

\begin{figure}
\begin{center}
	\subfigure[In-band alteration of packet by a middle-box. Only a single packet arrives at the client.]{
		\label{fig:inpath} 
		\includegraphics[width=0.4\textwidth]{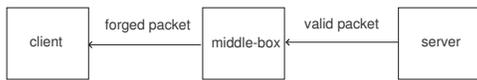}}
	\subfigure[Out-of-band injection of a forged packet. Two packets arrive at the client.]{
		\label{fig:outofband} 
		\includegraphics[width=0.4\textwidth]{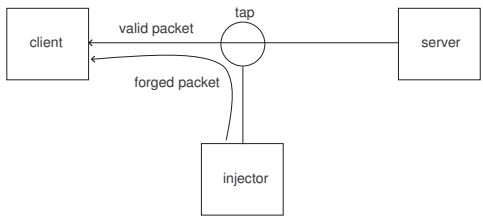}}
	\caption{In-band versus out-of-band alteration of content}
	\label{fig:alteration} 
\end{center}
\vspace{-10pt}
\end{figure}

Our analysis is based on the observation that network operators alter packets \emph{out-of-band}: all traffic is passively monitored, and when the content of a packet needs to be altered, a forged packet is injected into the connection between the server and the client. The forged packet poses as the valid packet. If the forged packet arrives at the client before the valid one, the client will accept the forged packet and discard the valid one. Such an approach has considerable advantages to the network operators since it does not introduce new points of failure to their traffic processing and there is no potential for a performance bottleneck.  Figure~\ref{fig:alteration} illustrates the differences between in-band alteration of traffic and out-of-band alteration. Note that both in-band and out-of-band traffic alteration is possible only on unprotected traffic, e.g., traffic that is not carried by TLS \cite{RFC5246} or authenticated using TCP authentication \cite{RFC5925}.

The out-of-band operation has a crucial characteristic that enables our analysis: the client receives two packets -- the forged one and the valid one -- that claim to be the same response from the server. However, they carry different content. This characteristic allows us to detect traffic alteration events while monitoring the traffic at the edge network. We can thus monitor and analyze traffic in a client-centric manner in which the traffic is not destined to a specific set of servers but to all servers contacted by the users at the edge network. Figure~\ref{fig:clientcentric} illustrates the traffic monitored in our work. In this paper we specifically focus our analysis on alteration of \emph{web} traffic, i.e., HTTP traffic over port 80.

\paragraph*{An example of out-of-band injection} To illustrate how content is altered using out-of-band injection, we describe in the following one of the injections we identified during our observations. In this example the user's browser sends the following HTTP GET request to cnzz.com (a Chinese company that collects users' statistics):
\small
\begin{verbatim}
GET /core.php?show=pic&t=z HTTP/1.1
User-Agent: Mozilla/5.0 (Windows NT 6.1; WOW64)
Host: c.cnzz.com
Accept-Encoding: gzip
Referer: http://tfkp.com/
\end{verbatim}
\normalsize
In response the user receives two TCP segments having the same value in the sequence number field. The segments include different HTTP responses. One segment carries the legitimate HTTP response that includes the requested resource (a JavaScript code) from cnzz.com:
\small
\begin{verbatim}
HTTP/1.1 200 OK
Server: Tengine
Content-Type: application/javascript
Content-Length: 762
Connection: keep-alive
Date: Tue, 07 Jul 2015 04:54:08 GMT
Last-Modified: Tue, 07 Jul 2015 04:54:08 GMT
Expires: Tue, 07 Jul 2015 05:09:08 GMT

!function(){var p,q,r,a=encodeURIComponent,c=...
\end{verbatim}
\normalsize

The other segment includes a forged response that directs the user via a 302 status code to a different URL that points to a different JavaScript code:
\small
\begin{verbatim}
HTTP/1.1 302 Found
Connection: close
Content-Length: 0
Location: http://adcpc.899j.com/google/google.js
\end{verbatim}
\normalsize
Our analysis shows that this JavaScript redirects the user through a series of affiliate ad networks ending with Google's ad network, which serves the user an ad. In this injection event the forged segment arrived before the legitimate one, which means that the user sees the injected ad instead of the original content.

\paragraph*{Relation to censorship} Website-targeted false content injection is similar in some ways to content blocking for the purpose of state-sponsored censorship. There is a substantial body of work that studies the mechanisms and characteristics of censorship worldwide~\cite{verkamp2012inferring, xu2011internet,levis2012collateral, aryan2013internet}. In many cases this blocking of content is also website-targeted. Moreover, blocking is often done by injecting false traffic segments, which in some cases is done out-of-band~\cite{verkamp2012inferring, clayton2006ignoring, towardsa}. In contrast to previous works on censorship, in this work we study the practice of false content injection by commercial network operators, rather than state entities. Such injections primarily serve financial gains rather than political agenda, with the goal of \emph{altering} the web content rather than blocking it. In this work we study and analyze the practice of financially-motivated false content injection by network operators. In Section~\ref{sec:relatedwork} we discuss in more detail related work on censorship. During this work we observed numerous occurrences of censorship-aimed injections. We do not report on them in this paper.

Our contributions can be summarized as follows:
\begin{enumerate}
	\item The observation that network operators inject false web content out-of-band.
	\item Investigation of the identities of network operators that practice website-targeted content injection.
	\item Thorough analysis of the characteristics of the injections and the purpose of the injecting operators. 
\end{enumerate}

The paper's structure is as follows. In Section~\ref{sec:background} we present technical background pertaining to injection of forged TCP and HTTP packets. Section~\ref{sec:methodology} details our methodology for monitoring web traffic and identifying injections of forged packets. Section~\ref{sec:datasources} details the sources of traffic we monitored. In Section~\ref{sec:analysis} we present our analysis of the injection events and our investigation as to the identities of the network operators behind them. Section~\ref{sec:mitigations} proposes effective and efficient client-side mitigation measures. Section~\ref{sec:relatedwork} discusses related work and Section~\ref{sec:conclusions} concludes the paper.

\section{Background} \label{sec:background}

\subsection{Out-of-band TCP Injection} \label{sec:TCPinjection}
A TCP \cite{RFC0793} connection between two end nodes offers reliable and ordered delivery of byte streams. To facilitate this service, every sent byte is designated a sequence number. Each TCP segment carries a Sequence Number field that indicates the sequence number of the first data byte carried by the segment. The following data bytes in the segment are numbered consecutively. A third party that wishes to send a forged TCP segment as part of an existing TCP connection must correctly set the connection's 4-tuple in the IP and TCP header, i.e., the source's port number and IP address as well as those of the destination. In addition, for the forged segment to be fully accepted by the receiver, the sequence numbers of the forged data bytes must fully reside within the receiver's  TCP window. 
Forging such a TCP segment is trivial for an on-path third party, since it can eavesdrop on the valid segments of the connection and discover the 4-tuple of the connection as well as the valid sequence number. 

In some circumstances an injected TCP segment may trigger an undesirable ``Ack storm". An ``Ack storm" occurs when the injected segment causes the receiver to send an acknowledgment for data bytes having sequence numbers  that were not yet sent by the peer. Appendix~\ref{ackstorm} details how an ``Ack storm" is formed. Nonetheless, as long as the injecting third party ensures that the injected TCP segment is no larger than the valid TCP segment sent by the peer, no ``ACK storm" will be triggered. If this is not the case, the injector could send a TCP reset right after the injection in order to forcibly close the connection. This will also eliminate the possibility of an ``Ack storm". The latter option is used only if the connection is expected to close right after the valid response is received. Indeed, in all our observations either of these alternatives took place and no ``Ack storms" were observed.

Nonetheless, the fact that the injected TCP segment aims to displace an already sent or soon to be sent valid TCP segment poses a different obstacle for the injecting third party.  According to the TCP specification \cite{RFC0793}, the first data byte received for a given sequence number is accepted. A subsequent data byte having  the same sequence number is always discarded as a duplicate regardless of its value. Thus, the injected segment must arrive at the receiver \emph{before} the valid TCP segment in order to be accepted. Note that  the TCP specification does not consider the receipt of bytes with duplicate sequence numbers as an error but rather as a superfluous retransmission. 

\subsection{HTTP Injection}
In this work we focus in particular on the injection of false HTTP responses received by a web client. HTTP \cite{RFC7230} is a stateless client-server protocol that uses TCP as its transport. An HTTP exchange begins by a client sending an HTTP request, usually to retrieve a resource indicated by a URI included in the request. After processing the request the server sends an HTTP response with a status code. The status codes we later refer to in this paper are: 
\begin{itemize}
	\item 200 (Successful): The request was successfully received, understood, and accepted. Responses of this type will usually contain the requested resource.
	\item 302 (Redirection): The requested resource resides temporarily under a different URI. Responses of this type include a Location header field containing the different URI. 
\end{itemize}

An HTTP client will receive only one HTTP response for a given request even when a false HTTP response is injected because, as mentioned above, the TCP layer will only accept the first segment that it receives (be it the false or the valid segment). When the forged response is shorter than and arrived before the valid response, the client then receives the byte stream that includes the forged response, followed by the tail of the valid response. The tail includes the data bytes having sequence numbers that immediately follow those of the forged response. By default, the response message body length is determined by the number of bytes received until the TCP connection is closed. This might be a problem for the injecting entity as the client will eventually receive a mixed HTTP response, which might yield unintended consequences. To avoid this problem, the injected response will usually include Content-Length or Transfer-Encoding headers that explicitly determine the end of the response. Thus, even if the TCP layer delivers the tail of the valid response to the HTTP layer, it will not be processed by the client.


\section{Methodology} \label{sec:methodology}
We now describe our methodology for collection and identification of TCP injection events. 
\begin{figure}
    \centering
    \includegraphics[width=0.5\textwidth]{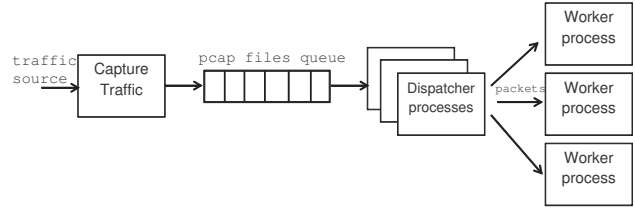}
    \caption{Depiction of the design of the monitoring system}
    \label{fig:monitor}
\vspace{-10pt}
\end{figure}

\subsection{Monitoring System}
At the core of the collection of  injection events was a monitoring system that eavesdropped on Internet traffic and identified these events. The monitoring system was deployed at the entry points of large networks (detailed in Section~\ref{sec:datasources}) and analyzed the bidirectional traffic that flowed in and out of those networks. The monitoring system was comprised of the following three stages (depicted in Figure~\ref{fig:monitor}). First, we captured the traffic using the 'netsniff-ng' tool \cite{netsniff-ng} along with a Berkeley packet filter \cite{mccanne1993bsd} to capture only HTTP traffic. The tool iteratively produced files comprising 200,000 packets each. These files were fed into a queue for processing by the next stage. To avoid explosion of the queue when the traffic rate exceeded the throughput of the next stages, the queue's length was bounded. Once the queue reached its limit, the capturing process was halted until the queue length  decreased.

At the next stage each capture file was processed by a dispatcher process that read each packet in the file, removed the Ethernet header, and computed a hash on the IP addresses and TCP ports in such a way that packets of the same TCP session would have the same hash result. A packet's hash result was then used to choose one of several worker processes to handle that packet. In this way all packets of the same session were delivered to the same worker.

At the final stage each worker process grouped the packets it received into TCP sessions and stored each session in a data structure. For each received packet a worker checked all the packets of that session to determine whether the conditions for a packet race were met (the conditions are detailed in Section~\ref{sec:injdetect}). If so, the last 30 packets of the session were written to a file, including their payload, for later analysis. See Section~\ref{sec:privacy} for the ethics and privacy issues pertaining to the storage and analysis of packets.
	
The packet sessions were stored by each worker in a data structure that is a least-recently-used cache with a fixed size.  Once the cache reached the maximum number of sessions it can store, the session that was idle the longest was evicted from the cache. To simplify packet processing we did not use TCP signaling (SYN and FIN flags) to create a new session in the cache or evict an existing one. This design choice gave rise to the possibility that a session would be evicted even if still active. Nonetheless, as our experiments show, the caches were large enough so that the minimum idle time after which a session was evicted did not drop below 10 minutes --- long enough to make the occurrences of active session evictions negligible. Note that even if such an eviction were to occur, packet races could still be detected  in that session, since we treated the packets sent after the idle period as a new session and stored them in the cache. In this case, however, the packets of the session prior to the eviction would not be available for analysis.

\subsection{Injection Detection} \label{sec:injdetect}   
The detection logic of packet injection events is relatively straightforward. Our goal was to detect packet races within the session, namely, two packets that carry different payloads, but correspond to the same TCP sequence numbers. Usually these packets will arrive in quick succession. To make our code more efficient we checked for a race only between  pairs of packets that were received within a time interval that does not exceed the parameter $MaxIntervalTime$. Throughout our data collection process we set $MaxIntervalTime = 200msec$. We believe that this value captures the vast majority of  injection events as almost all round trip times on the Internet are below 400msec \cite{rtt}. Indeed, nearly all of the time differences we observed between raced packets were below 100msec (see Section~\ref{sec:analysis}). Algorithm~\ref{alg:racedetection} in Appendix~\ref{sec:injectdetectalg} details the procedure for race detection. 

The procedure we used to identify packet races should, in theory, flag only events in which a third party injected rogue packets into the TCP session. However, interestingly, we observed numerous events which fulfill the above conditions but are not the result of a packet injection. We detail such occurrences in Appendix~\ref{sec:falsepositives}.

\begin{table*}[t]
	\centering
		\begin{tabular}{|C{3cm}||C{2cm}|C{2cm}|C{2cm}|C{3cm}|}
			\hline
			Institution & User base & Monitoring period [week] & Traffic volume [Tb] & Number of sessions [Million] \\
			\hline
			University A & 20,000 & 2 & 80 & 8 \\
			\hline
			University B \& University C & 50,000 & 16 & 1400 & 120 \\
			\hline
			Enterprise D & 5,000 & 3 & 24 & 0.8 \\
			\hline
		\end{tabular}
		\vspace{12pt}
	\caption{Monitored traffic sources}
	\label{tab:MonitoredTrafficSources}
	\vspace{-10pt}
\end{table*}

\subsection{Ethics and Privacy} \label{sec:privacy}
As explained above, the monitoring system captures Internet user traffic. To minimize concerns about user privacy, the system stores only TCP sessions in which a packet race was detected. All other sessions are only cached briefly in the workers' caches, after which they are permanently erased. Moreover, for each stored session, only the last 30 packets (at most) are saved. Earlier packets are dropped. This is in order to store only those packets that are relevant to the analysis of the injection events while minimizing the chance that user privacy will be breached. Indeed, during our analysis no identifiable personal information was found in the stored sessions.

Throughout our research we were supervised by the networks' administration teams, who reviewed and approved the code of the monitoring system and procedures for the analysis of the stored sessions. During the analysis the location and identity of users associated with IP addresses were never disclosed to us. Finally, we note that our monitoring system passively collected information; it never interfered or tampered in any way with the traffic.

\subsection{Limitations}
Our monitoring system cannot detect content alterations in which there is no race between the legitimate packet and the forged one. In particular, we cannot detect the following cases:
\begin{enumerate}
	\item In-band changes in which the legitimate packet is changed in-place. In such cases the client only sees a forged packet.
	\item Additions to the response in which an extra forged packet is sent such that it extends the HTTP response, but does not replace any legitimate part.
	\item Drops of packets that are part of a valid HTTP response.
\end{enumerate}

We monitored a large volume of traffic originating from diverse networks having tens of thousands of users (see Section~\ref{sec:datasources}). Nonetheless, as in any other study that involves uncontrolled traffic, our findings are only as diverse as the traffic we monitor. Namely, we cannot identify an injecting entity on the Internet if we do not monitor traffic that triggers an injection by that entity. Furthermore, the types of injections we have observed are dependent on the web traffic originating from the networks we monitored.

\section{Data Sources} \label{sec:datasources}
During our study we monitored the network traffic of four institutions. For each institution we monitored the Internet traffic (incoming and outgoing) of all its users. In all cases the same monitoring mechanism was used: traffic was copied to the monitoring system using a SPAN port out of a border switch. In all cases, we only monitored HTTP traffic, namely traffic having source port or destination port that equals 80. 

Table~\ref{tab:MonitoredTrafficSources} lists the  characteristics of the monitored traffic sources. For each institution we list the number of users who may use Internet connectivity in that institution. For a university this is the number of students and staff, and for an enterprise this is the number of employees. In addition, we list the length of time we monitored the traffic as well as the total volume traffic and number of sessions the monitoring system processed. In aggregate, we monitored the traffic of more than 75,000 users, while processing 1.4 petabits carried by 129 million HTTP sessions contacting servers having more than 1.5 million distinct IP addresses. The details of University B and C are displayed together since we monitored their traffic jointly on the same border switch.  Enterprise D represents the main branch of a large hi-tech company. The monitored branch includes an extensive R\&D division as well as the headquarter offices and the international marketing and sales divisions. All institutions wish to remain anonymous.

\section{Injection Analysis} \label{sec:analysis}
In this section we present an analysis of the injection events. In Section~\ref{sec:InitialInvestigation} we present an overview of the injections and highlight a few of them. Section~\ref{sec:tellingapart} describes ways to automatically distinguish between the valid and forged packets. In Section~\ref{sec:timing} we explore the time differences between the raced packets. Section~\ref{sec:repeatability} characterizes the recurrence of injection events. Finally, Section~\ref{sec:whoisbehind} presents an investigation aimed at unveiling the entities behind the injection events.

\subsection{Initial Investigation} \label{sec:InitialInvestigation}
In this section we refer to a TCP session into which a forged packet was injected as an \emph{injected session}.  We manually analyzed each injection event. We detected around 400 injection events that aim to alter web content\footnote{We have also found hundreds of additional events that do not aim to alter web content; these events were related to caching and censorship.}. Although this is not a negligible number, it pales in comparison to the total volume of traffic we monitored to extract these events. This is attributed to the fact that most of the injected sessions were destined to web servers in the Far East, a region to which relatively little traffic is destined from the networks we monitored. Thus the relatively small number of injections. Nonetheless, these events were sufficient to gain substantial indications as to the different entities that practice forged content injection (Section~\ref{sec:whoisbehind}). 

We grouped the injection events into 14 groups based on the resource that was injected into the TCP session. In other words, two injections that forged the same content are placed in the same group. Representative (and anonymized) captures of the injected sessions can be found in \cite{publishinjections}. For each injection group we publish up to 4 captures of injected sessions that are representative of their respective group. To preserve the anonymity of the users, in each capture we zeroed the client's IP address as well as the IP and TCP checksum fields.

Table~\ref{tbl:injections} lists the groups. For each group we list the following details:
\begin{enumerate}
	\item Group name -- an identifier that was given by us to that group. We selected the name either by the name of the site whose content was forged or by the name of a server the forged content directed us to.
	\item Destination site(s) -- the website(s) of the requested resource that was forged. There may be several such sites for a single group.
	\item Site type -- the category of the destination site(s)
	\item Location -- the country in which the destination site's server resides
	\item Injected resource -- the type of forged content that was injected
	\item Purpose -- the aim of the injection
\end{enumerate}
 
\begin{table*}
	\centering
		\begin{tabular}{|c|c|C{2cm}|c|C{3.5cm}|C{2cm}|}
			\hline
Group name	& Destination site(s) &	Site type	& Location	\comment{&  injections	\#} & Injected resource &	Purpose \\ \hline
szzhengan	& wa.kuwo.cn	& Ad network	& China \comment{& 	1}	& A JavaScript that appends content to the original site & Malware \\ \hline
taobao	& is.alicdn.com	& Ad network &	China	\comment{& 7}	& A JavaScript that generates a pop-up frame & Advertisement \\ \hline
netsweeper	& skyscnr.com	& Travel search engine & India \comment{& 2} 		& A 302 (Moved) HTTP response & content filtering \\ \hline
uyan	& uyan.cc	& Social network & China \comment{&	2}	 & A redirection using 'meta-refresh' tag & Advertisement \\ \hline
icourses	& icourses.cn	& Online courses portal	& China \comment{&	1}	& A redirection using 'meta-refresh' tag & Advertisement \\ \hline
uvclick	 & cnzz.com &	Web users' statistics & Malaysia \comment{& 21}	& A JavaScript that identifies the client's device & Advertisement \\ \hline
adcpc	 & cnzz.com	&		Web users' statistics  & Malaysia \comment{& 4}	& A 302 redirection to a JavaScript that opens a new window & Advertisement \\ \hline
jiathis &	jiathis.com	& Social network &	China \comment{& 5}	& A redirection using 'meta-refresh' tag & Advertisement \\ \hline
server erased	 & changsha.cn & Travel & China \comment{&	16}	& Same as legitimate response but the value of HTTP header 'Server' is changed & Content filtering \\ \hline
GPWA	& gpwa.org	& Gambling & United States	\comment{& 2}	& A JavaScript that redirects to a resource at qpwa.org & Malware \\ \hline 
tupian	& \begin{tabular}{@{}c@{}}www.feiniu.com \\ www.j1.com  \end{tabular} & e-commerce & China \comment{& 3} &	A JavaScript the directs to a resource at www.tupian6688.com	& Malware \\		\hline
mi-img & mi-img.com & Unknown & China \comment{& 2} & A 302 redirection to a different IP & Malware \\ \hline
duba & unknown & Unknown & China \comment{& 5} & A JavaScript that prompts the user to download an executable & Malware \\ \hline
hao & 02995.com & Adware-related & China \comment{& 1} & A 302 (Moved) HTTP response & Advertisement \\ \hline
		\end{tabular}
	\caption{Injection groups and their characteristics}
	\label{tbl:injections}
\end{table*}

It is evident from Table~\ref{tbl:injections} that the majority of injected sessions we observed were to web servers located in China. We note that the networks we monitored are \emph{not} located in China or the Far East, but in a Western country. The proportion of HTTP traffic destined to China in the monitored networks is only about 2\%.  This is a first indication that the majority of entities that injected the forged content we observed reside in China (we investigate the injectors' identity in Section~\ref{sec:whoisbehind}).

Seven injection groups are aimed at injecting advertisements to web pages. An analysis of the injected resources shows similarities between the various groups. These similarities might indicate that the injections are done by the same entity or at least by different entities that use the same injection mechanism or product. The injection groups 'icourses', 'uyan', and 'jiathis' all used the HTML meta refresh tag to redirect the user to a different URL. In all cases, the redirection was to Baidu (a Chinese search engine) using the URL \url{www.baidu.com/?tn=95112007_hao_pg}. The URL includes a referral tag that identifies \url{hao123.com} -- a well-known adware-related site -- as the referring site. The referral tag is possibly used by Baidu to pay hao123 for referring traffic to it. In one case, the redirected URL included a search keyword for a clothing chain store. Interestingly, another injection group, 'hao', referred the user to \url{hao123.com} itself, but using a different mechanism -- an HTTP 302 response.

Surprisingly, five injection groups showed strong indications that the aim of the injector was malicious. One such group is 'gpwa'. The injections in this group target the traffic to \url{gpwa.org}. The forged content here includes a JavaScript that refers to a resource having the same name as the one originally requested by the user, but the forged resource is located at \url{qpwa.org}, a domain that is suspiciously similar to the legitimate domain. The forged domain is registered to a Romanian citizen, who appears to be unrelated to the organization that registered the domain \url{gpwa.org}. These are strong indications of malicious intent. The site \url{qpwa.org} was already brought down by the time we analyzed this injection group.

The injections in the 'duba' group add to the original content of a website a colorful button that prompts the user to download an executable from a URL at the domain \url{duba.net}. The executable is flagged as malicious by several anti-virus vendors.  

Another malicious injection group is 'mi-img'. In these injected sessions the client, which appears to be an Android device, tries to download an application. The injected response is a 302 redirection to another IP address (no domain name is specified). According to BotScout~\cite{botscout} -- an online bot database -- this forged IP address is known to be a bot. We retrieved the application from this IP address. The downloaded apk file is flagged by Fortinet's antivirus as a malware called 'Android/Gepew.A!tr'.

Another injection group worth mentioning is 'server erased'. In this group injections were identical to the legitimate response but instead of original value of the Server HTTP header, e.g., nginx/1.2.7, the string '*******' appeared. This is as if to prevent identification of the web server's software. The motivation for this injection is unclear.

\subsection{Distinguishing the Forged Response from the Valid One} \label{sec:tellingapart}
Identifying a race between two packets is a relatively straightforward task. However, without a priori knowledge of the legitimate content expected from the server, automatically distinguishing the forged packet from the legitimate one is not trivial.  Nonetheless, in the following we list a few rules that worked well for this difficult task.

\paragraph*{IP identification} In many operating systems, such as Windows and Linux \cite{herzberg2012security}, the IP identification value equals a counter that is incremented sequentially with each sent packet. Is some operating systems there is a single global counter for all sessions. In others, there is a separate counter for each destination. Indeed, our observations show that in most injected sessions the IP identification values of the packets sent by the web server are either monotonically increasing (when the counter is global) or consecutively increasing (when there is a counter per destination). In most of the injection events we observed that the injecting entity made no attempt to make the identification value of the forged packet similar to the identification values of the other packets sent by the server. In Appendix~\ref{sec:injectedid} we detail a few of the (failed) attempts of the injecting entity to mimic the Identification field of the legitimate packet it aims to displace.

We formulate the following rule to determine which of the two raced packets is the forged one: the forged packet is the one that has the largest absolute difference between its identification value and the average of the identification values of all the other packets (except the raced one).

For all injection events, we manually identified the forged packet according to its content and compared it to the corresponding identification that used the above rule. The comparison reveals that the rule is accurate about 90\% of the time. This is a fairly accurate measure considering that it is not based on the payload of the raced packets. 

\paragraph*{IP TTL} The IP TTL value in a received packet is dependent on the initial value set by the sender and the number of hops the packet has traversed so far. Thus, it is unusual for packets of the same session to arrive at the client with different TTL values. Therefore, if the raced packets have different TTL values we can use them to distinguish between the two packets. From our observations, the injecting entity often made no attempt to make the TTL value of the forged packet similar to the TTL values of the other packets sent by the server. Similarly to the case of the IP identification rule above, we identify the forged packet using the following rule: the forged packet is the one that has the largest absolute difference between its TTL value and the average of TTL values of all the other packets (except the raced one). 

Manual analysis of the injection events reveals that the TTL rule correctly identified the forged packet in 87\% of all injection events. The TTL rule concurs with the IP identification rule above in 84\% of all injection events. We thus conclude that the TTL and identification values can serve to effectively distinguish the forged packet from the valid packet.

We note that our finding that the TTL and Identification fields of the forged packets have abnormal values generally agrees with findings on censorship-related injections which also show that censoring entities do not align the TTL and Identification values with those of the legitimate packets (e.g., \cite{towardsa}). 



\subsection{Timing Analysis} \label{sec:timing}
\begin{figure}
    \centering
    \includegraphics[width=0.5\textwidth]{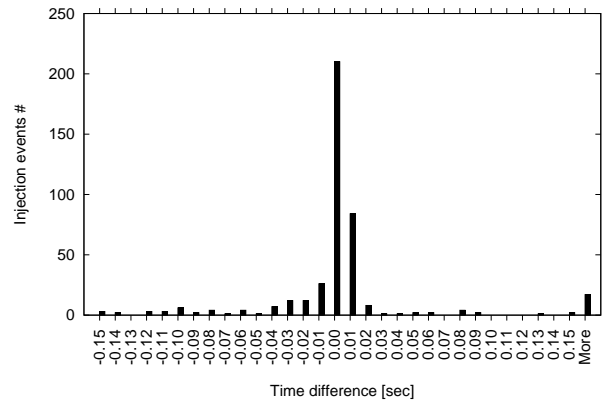}
    \caption{Arrival time difference between the forged and legitimate packets}
    \label{fig:timehist}
\end{figure}

The race between the forged and legitimate packets can also be characterized by the difference in their arrival times. By arrival time we mean the time at which the packet was captured by the monitoring system. Since the system captures traffic at the entrance to the edge network close to the client, it is reasonable to assume that these times are very close to the actual arrival times at the end client. For each injection event we calculate the difference between the arrival time of the legitimate packet and the arrival time of the forged packet. A negative difference means that the forged packet ``won" the race, and a positive difference means that the legitimate packet ``won". The histogram of the time differences of all the injection events we observed are shown in Figure~\ref{fig:timehist}.

It is evident from Figure~\ref{fig:timehist} that in most injection events the forged packet wins the race. In only 32\% of the events does the legitimate packet arrive first. This result strengthens our initial assumption that the decision to inject a forged packet is made according to the HTTP request sent by the client. This means that the injecting entity can send the forged packet well  before the server sends the legitimate packet, as the client's request still needs to travel to the server. Still, even in such a case, in a non-negligible portion of events, the forged packet loses the race. This may indicate injections that occurred very close the server. Alternatively, it may indicate that in some cases the decision to inject the packet is made at the time the response from the server is encountered. In the latter case, the forged packet is at a distinct disadvantage as it starts the race lagging behind the legitimate packet. In many cases in which the forged packet  won the race, the legitimate packet arrived very soon after, in less than 10msec. 


\subsection{Repeatability} \label{sec:repeatability}
All injection groups were observed for only a short period of time, usually one to three days, after which they were not detected again by our monitoring system. A few injection types were even encountered only once. No long-term (3 days or more) injections were observed by our monitoring system\footnote{The only long-term injections we did observe were related to censorship and caching. These injections were the only ones we were able to reproduce. }. 

We next tried to reproduce the injection events we observed. For each injection event we extracted the HTTP request that triggered the injection. We then sent from the edge network in which the injection originally occurred the same HTTP request. We captured the resulting TCP sessions and searched for injections. We were not able to reproduce any of the injection groups.

From the above findings we surmise that, in general, injections by on-path entities may be intermittent; namely, the injecting entity injects forged content to a particular site for only a short period of time before moving on to other sites. This might be motivated by the desire of the injector to stay ``under the radar". It is plausible that injecting forged content to a site for only a short period of time might go unnoticed by the users and site owners, or at least would not cause them to expend effort investigating the forged content's origin.


\subsection{Who is Behind the Injections?} \label{sec:whoisbehind}
We finally turn our attention to the culprits behind these injection events. In general, it is  difficult to unveil these entities as there is no identifying information in the injected content. Nonetheless, we can get indications as to the identity of the injecting entities by trying to detect the autonomous system from which the forged packet originated. We assume that the entity that operates this autonomous system is the entity responsible for the injection. 

We note that since the injections were not reproducible during this analysis, we cannot employ the oft-used traceroute-like procedure to locate the injector \cite{levis2012collateral,towardsa,marczak2015analysis}. In this procedure the packet triggering the injection is repeatedly sent with increasing TTL values until the forged response is triggered, thereby revealing the location of the injector. To identify the injecting entities we resort to the following procedure:
\begin{enumerate}
	\item Estimate the number of hops the forged packet traversed: this estimation relies on the packet's TTL value. Specifically, it relies on there being a significant difference between the default initial TTL values set by the major operating systems \cite{DefaultTTLs}: in general, the differences between those initial values are larger than the length of most routes on the Internet. The default initial TTL values of the major operating systems are 32, 64, 128 and 255. This means, for example, that if a packet is received with a TTL value of 57, the initial TTL value of that packet was likely to be 64 and the number of hops traversed was likely to be 7. If the estimated number of hops is larger than 30 or smaller than 3 \footnote{Nearly all routes on the Internet are shorter than 30 hops \cite{leguay2005describing}. Additionally, it is very unlikely that the injecting third party resides less than 3 hops away since the first couple of hops reside within the edge networks we were monitoring.}, we assume the estimation is incorrect and stop the analysis.
	\item Identify the path from the destination server to the client: the actual path from the server to the client cannot be known without an agent  in the server's network. Instead, we use the path from the client to the server while assuming that the routing on this path is symmetric. We identify the path from the client to the server by using a 'traceroute' tool. The traceroute used a TCP syn packet with destination port 80. We found that such a packet triggers responses from most routers and servers. 
	\item Infer the hop along the above path from which the forged packet was injected: using the estimated number of hops the forged packet traversed and the estimated path it traversed, we can now infer the hop on the path from which the packet was sent. 
	\item Identify the autonomous system the injecting hop belongs to: given the IP address of the hop, we can now identify the autonomous system to which it belongs in order to reveal the entity responsible for injecting the packet. To this end we leveraged public databases that hold current BGP advertisements: this allows us to identify the autonomous system that advertises the given IP address. BGP advertisements for mapping of IP addresses to autonomous systems are known to be more precise and up-to-date than Internet route registries \cite{mao2003towards}. 
\end{enumerate}
It should be noted that this procedure has the following caveats:
\begin{enumerate}
	\item The initial TTL value of the injected packet may not be one of the common default values. In such cases, this analysis can not be carried out. In particular, based on the TTL values of the injected packet, we conclude that this is indeed the case for the injections in the groups 'jiathis', 'uyan', 'mi-img', and 'icourses'.
	\item Not all routes on the Internet are symmetric. If the path from the client to the server is not symmetric, the analysis will produce an incorrect result. We address this issue in the next subsection.
	\item The implicit assumption of this procedure is that the injecting machine resides on-path. Strictly speaking, this need not be the case. An on-path machine monitoring the traffic can trigger the injection from a remote machine. In such a case the forged packet will travel on an entirely different path than the legitimate packets.
\end{enumerate}
 
\begin{table}[t]
	\centering
		\begin{tabular}{|c|C{2cm}|C{2cm}|}
			\hline
			Injection group & Web server's AS number & Suspected injecting AS number\\
			\hline
			xunlei & 17816 &  17816\\
			\hline
			szzhengan & 4134 &  4134  \\
			\hline
			taobao & 4837 & 4837 \\ \hline
			uvclick & 38182 &  38182 \\ \hline
			adcpc & 38182  & 38182\\ \hline
			server erased & 4134 & 4134 \\ \hline
			GPWA & 6943 & 6943 \\ \hline
			tupian & 4812 & 4812 \\
			\hline
		\end{tabular}
		\vspace{12pt}
	\caption{The autonomous system numbers in which the injected web servers reside and in which the suspected injecting entities reside}
	\label{tbl:InjectorIdentities}
\end{table}

In Table~\ref{tbl:InjectorIdentities} we list the results of the above analysis. For each injection group, we list the autonomous systems in which the destination sites reside and the autonomous systems suspected of the injections. The table lists only injection groups for which the analysis can be performed; namely, the estimated number of hops the injected packet traversed is not larger than 30 and not smaller than 3, and it is also not larger than the path between the client and server.

It is evident from Table~\ref{tbl:InjectorIdentities} that in all cases where the above analysis succeeded, the forged content was injected in the very same autonomous system where the destination site resides. This is not to say that the injection was done by the web server itself or by the entity responsible for its content. In many cases the hop suspected to be the one the injection was performed on is not the last hop on the route to the server. The most reasonable explanation for this phenomenon is that the network operators who inject the forged content do so without the knowledge of the web servers they host or provide Internet access to. 

Table~\ref{tbl:ASoweners} lists for each suspected injecting autonomous system the organization that operates it. It is worth noting that two of the largest network operators in China -- China Unicom and China Telecom -- are suspected of practicing content injections. Moreover, the autonomous systems of these operators originate injections of different groups. This might imply that more than one injector mechanism is deployed in these autonomous systems. 

The operator of the suspected autonomous system for the 'gpwa' group is Information Technology Systems. In this particular case, this is the organization that is responsible for the content of the destination site for these injections -- \url{gpwa.org}. Since there are strong indications that the injections of this group are malicious (see discussion in Section~\ref{sec:InitialInvestigation}), we assume that the attacker compromised a router in the suspected autonomous system.

\begin{table}[t]
	\centering
		\begin{tabular}{|c|c|}
			\hline
			AS number & Operator\\
			\hline
			17816, 4837 & China Unicom \\
			\hline
			4134, 4812 &  China Telecom \\
			\hline
			38182 &  Extreme Broadband (Malaysia) \\ \hline
			6943 &  Information Technology Systems (US)\\ \hline
		\end{tabular}
		\vspace{12pt}
	\caption{The operators for each suspected injecting autonomous system}
	\label{tbl:ASoweners}
\end{table}

\vspace{5pt}

\subsubsection*{Using a traceroute from the server-side}
As noted above, a caveat of the above analysis is that we used traceroutes from the client to the server while assuming this route is symmetric. This is a necessity since we cannot execute a traceroute to the client from the actual server. To address this caveat we leveraged RIPE Atlas~\cite{RIPEAtlas}. This is a global network comprised of thousands of probes hosted throughout the Internet. Each probe can be instructed to execute a measurement out of a predetermined set that includes ping, DNS query, HTTP request and traceroute. RIPE Atlas hosts 6 probes in 3 autonomous systems that host destination sites the content of which was forged. The autonomous systems and the corresponding injection groups that forged content for destination sites residing in each autonomous system are listed in Table ~\ref{tbl:RIPEAS}.

\begin{table}[t]
	\centering
		\begin{tabular}{|c|c|}
			\hline
			AS number & Injection groups \\
			\hline
			4812 & tupian\\
			\hline
			4134 & szzhengan, server erased \\
			\hline
			4808 & uyan, icourses, jiathis \\
			\hline
		\end{tabular}
		\vspace{12pt}
	\caption{Autonomous systems which host RIPE Atlas probes and the corresponding injection groups that forged traffic of web servers residing in those autonomous systems}
	\label{tbl:RIPEAS}
\end{table}

For each of the 6 probes we executed a traceroute from it to the edge network where the corresponding injection events were identified. We then employed the procedure we described above on these new traceroutes. We note that using these traceroutes may still not be without error. The probes indeed reside in the autonomous systems that host the destination site; however, we cannot guarantee that their route to the client is the same as the route from the destination site. Specifically, the traffic from the probe may exit the autonomous system through a different point than the traffic originated from the site. 

The traceroute from each of the 3 autonomous systems to the corresponding edge network were \emph{different} than the opposite routes from the edge networks to those autonomous systems. Nonetheless, in all cases, a pair of routes in opposite directions traversed the same autonomous systems with the exception of one Tier-1 autonomous system; namely, each route traversed a different Tier 1 operator (for example, the route between the client and the server traversed Level 3's AS while the route in the opposite direction traversed Cogentco's AS). The other autonomous systems on the routes were the same; this is why the outcome of the analysis with these routes was the same as for the routes in the opposite direction. The analysis for the 'szzhengan' and 'server erased' injections yielded the same suspected autonomous system -- 4134, while the analysis for the 'tupian' injections yielded a different autonomous system -- 4134 instead of 4812 found by the previous analysis. Nonetheless, these autonomous systems are siblings operated by the same company -- China Telecom. 

The injecting groups that correspond to destination sites residing in autonomous system 4808 -- 'uyan', 'icourses', and 'jiathis' -- were set with an unknown initial TTL value (namely the estimated number of hops was larger than 30 or smaller than 3); hence the analysis cannot be performed on them.

\section{Proposed Mitigation} \label{sec:mitigations}
The best mitigation against TCP injection attacks is simply to use HTTPS. Unfortunately, this is not always subject to the discretion of the user. Many web sites still do not support HTTPS~\cite{httpsanalysis}. A user wishing to access a website that does not support HTTPS must resort to the unprotected HTTP. In this section we present a client-side mitigation measure that monitors the incoming HTTP traffic and blocks injected forged TCP segments, thereby defending the user even if he must use HTTP.

A naive mitigation measure is to simply apply the procedure described in Algorithm~\ref{alg:racedetection} on the monitored traffic in order to identify packet races. Nonetheless, such an approach means that every incoming packet must be delayed for 200msec. Such a delay is necessary in order to make sure a given packet is not an injected packet forging a legitimate one. Only after 200msec have passed with no race detected can we accept the packet. Such an approach incurs noticeable delay on the incoming traffic and degrades the user's browsing experience. This approach, however, ensures that all injected packets will be identified and blocked. In Section~\ref{sec:mitigation:exp} we detail our experimental results with such an approach. 

An improved approach is to take advantage of the insights we presented in Section~\ref{sec:tellingapart}, where we showed that for the vast majority of the injected packets, the values of the TTL and Identification fields in the IP header do not correspond to the respective values of the legitimate packets of the session. This insight can be leveraged to improve the naive mitigation measure such that only packets with abnormal TTL or Identification values will be delayed for 200msec, and only for those packets will we try to detect a race. This way only suspicious packets are delayed. 

Algorithm~\ref{alg:mitigation} in Appendix~\ref{sec:mitigationalg} details the improved mitigation algorithm. Note that this algorithm will be effective only if the forged packets exhibit anomalous TTL or Identification values as compared to the legitimate packets in the injected session. We note that it is possible for an injector to inject a packet with values that will not appear anomalous, as in most likelihood it can also inspect the traffic sent by the web server.  Anomalous TTL and identification values have also been observed in the censorship-related state-sponsored injections \cite{towardsa}. This indicates that aligning the TTL and identification values to the legitimate values might not be trivial to implement. Indeed, aligning the identification value requires that injector keep track of the identification values of packets sent by the web server for every potential injection session, well before the actual injection decision is made. This may require a substantial addition of memory space and computational overhead. If the injector does align the TTL and identification values, the improved mitigation algorithm we propose will not be not be effective and the naive approach must be used.

\subsection{Experiments} \label{sec:mitigation:exp}
We now detail our experiments to evaluate the two mitigation algorithms -- the naive and improved algorithms. We evaluate the algorithms using two measures:
\begin{enumerate}
	\item Web page load time increase -- this measure shows the increase of time it takes to load a web page as compared to the case where no mitigation measure is employed. This measures the extent to which the algorithm degrades the user's experience.
	\item False negatives -- this measure counts how many injections are not identified. This measures the effectiveness of the algorithm.
\end{enumerate}

We evaluated the algorithms against two data sets:
\begin{enumerate}
	\item Benign data set -- this data set includes traffic of benign web browsing having no content injection. We used the 200 most popular sites from Alexa's list~\cite{alexa}. From these sites we used the ones for which majority of their objects are fetched using HTTP (rather than HTTPS). There are 136 of these sites that met this criterion. 
	\item Injected data set -- This data set includes the injected sessions we captured throughout our observations.
\end{enumerate}
The two algorithms were evaluated on the benign data-set to measure the web page load time increase. We browsed each website using PhantomJS. We inspected the incoming traffic while leveraging the NFQUEUE target of Linux iptables~\cite{nfqueue}. We measured the load time of each website 5 times and recorded the smallest load time value to disregard intermittent network delays. We compared these load times to the load times where no mitigation algorithm is deployed.

The two algorithms were evaluated on the injected data set to measure the false negative events, i.e., the injections that were missed.  Table~\ref{tbl:mitigation:results} summarizes the findings. It is evident that the naive algorithm imposes a considerable increase in page load time -- 120\%. In contrast, the improved algorithm incurs a mere 12\% increase, while having a negligible false negative rate of 0.3\%.

\begin{table}[t]
	\centering
		\begin{tabular}{|c|C{2cm}|C{2cm}|}
			\hline
			Algorithm & Load time increase & False Negative\\
			\hline
			naive & 120\% &  0\%\\
			\hline
			improved & 12\% &  0.3\%  \\
			\hline
		\end{tabular}
	\caption{The performance of the two mitigation algorithms.}
	\label{tbl:mitigation:results}
\end{table}

\section{Related Work} \label{sec:relatedwork}
The practice of Internet traffic alteration has been studied in several works \cite{kreibich2010netalyzr, weaver2014here, reis2008detecting, Zimmerman2015}, all of which have employed the server-centric approach described in the Introduction. 

In~\cite{kreibich2010netalyzr, weaver2014here} the authors deployed a website that directs users to about 20 back-end servers that deliver a Java applet. The applet runs a series of tests which try to fetch predetermined content. The analysis found many web proxies of several categories, the most popular of which are anti-virus software installed on the end clients, HTTP caches and transcoders deployed by ISPs, and security and censor proxies deployed by enterprises and countries. Ref.~\cite{weaver2014here} identifies  two ISPs that employ HTTP error monetization, and one that injects advertisements into all HTTP connections.

In~\cite{reis2008detecting} the authors set up a web server that delivers the same content from a handful of different domains. The content includes a JavaScript code that runs when the page is loaded in the client's browser and reports any detected changes to the web page. It found that most changes to the content were made indiscriminately regardless of the originating domains. Most of the content modifications were due to software installed locally on the end clients or due to security gateways deployed at enterprises. Other modifications were due to ISPs that compressed content delivered to their users. Additionally, 4 ISPs and a company that provides
free wireless service were identified as injecting advertisements to web pages their customers visit.

In~\cite{Zimmerman2015} the authors leveraged the online advertising infrastructure of several ad networks to spread a specially crafted Flash-based advertisement that runs a JavaScript code and retrieves a preconfigured measurement page while reporting back any change made to it. Almost 1000 page alteration events were detected; however, the portion of events for which ISPs are responsible is unknown.  

The authors of~\cite{zhang2011inflight} investigate inflight modifications of traffic from an unnamed popular Internet search service. In contrast to the abovementioned works, here the changes were detected by the IP address the client contacted, which was different than  the addresses owned by the search service. This work found 9 ISPs that proxy their customers' traffic destined to the search service. The redirection to the proxy is done by resolving the DNS name of the service to the IP address of the proxy.

A considerable body of work deals with censoring countries and the mechanisms they use to censor Internet traffic. The authors of~\cite{verkamp2012inferring} have categorized the mechanisms of the censorship employed by different countries. It is noted that China and Thailand use out-of-band devices to send forged packets, which are usually HTTP 302 redirection, or a TCP reset.

In~\cite{weaver2009detecting} it was shown that several ISPs enforce usage restrictions of their networks by actively terminating undesirable TCP connections. The authors note that this is done by sending forged TCP resets out-of-band. They then leverage this insight -- much as we do in the current work -- to identify these forged resets. Nonetheless, the detection conditions are different than the ones we used since the forged TCP reset has no payload to spoof; hence, the detection conditions mainly revolve around the arrival time and sequence number of the reset segment as compared to those of other segments in the connection.

The authors of \cite{duan2012hold} discuss attacks that employ out-of-band injection of forged DNS responses. To mitigate the effects of such attacks it is suggested that the resolver wait after receiving an initial reply to allow a subsequent legitimate reply to also arrive. In particular, the resolver should wait for another reply if the first reply arrived sooner than half of the expected RTT since the query was issued or if the TTL field in the IP header does not have the expected value. If indeed two replies eventually arrive, this indicates an attack.

\section{Conclusions} \label{sec:conclusions}
In this work we reveal a new side to the practice of false content injection on the Internet. Previously, discussion on this practice focused on edge ISPs that limit their misdeeds to the traffic of their customers. However, we discovered that some network operators inject false content to the traffic of predetermined websites, regardless of the users that visit them. Our work leverages the observation that rogue content injection is done out-of-band. It can hence be identified while monitoring an edge network in which the victim clients reside. Our analysis is based on extensive monitoring of a large amount of Internet traffic. We reveal 14 groups of content injections that primarily aim to impose advertisements or even maliciously compromise the client. Most of the financially-motivated false content injection we observed originated form China. Our analysis found indications that numerous injections originated from networks operated by China Telecom and China Unicom -- two of the largest network operators in Asia.

\section*{Acknowledgments}
The authors would like to thank Hank Nussbacher and Eli Beker, whose cooperation made this research possible.

{\footnotesize \bibliographystyle{acm}
\bibliography{bibdb}}

\begin{thebibliography}{10}

\bibitem{alexa}
{Alexa}.
\newblock \url{http://www.alexa.com/}.

\bibitem{botscout}
{BotScout}.
\newblock \url{http://botscout.com/}.

\bibitem{netsniff-ng}
netsniff-ng toolkit.
\newblock \url{http://netsniff-ng.org}.

\bibitem{publishinjections}
Representative captures of the injected sessions.
\newblock
  \url{http://www.cs.technion.ac.il/~gnakibly/TCPInjections/samples.zip}.

\bibitem{httpsanalysis}
Ssl/tls analysis of the internet's top 1,000,000 websites.
\newblock \url{https://jve.linuxwall.info/blog/index.php?post/TLS_Survey}.

\bibitem{nfqueue}
{Using NFQUEUE and libnetfilter\_queue}.
\newblock
  \url{https://home.regit.org/netfilter-en/using-nfqueue-and-libnetfilter_queue/}.

\bibitem{CMA}
{\sc Anderson, N.}
\newblock {How a banner ad for H\&R Block appeared on apple.com}.
\newblock
  \url{http://arstechnica.com/tech-policy/2013/04/how-a-banner-ad-for-hs-ok/}.

\bibitem{towardsa}
{\sc Anonymous}.
\newblock Towards a comprehensive picture of the great firewall’s dns
  censorship anonymous.
\newblock In {\em 4th USENIX Workshop on Free and Open Communications on the
  Internet (FOCI 14)\/} (2014).

\bibitem{aryan2013internet}
{\sc Aryan, S., Aryan, H., and Halderman, J.~A.}
\newblock Internet censorship in iran: A first look.
\newblock In {\em Proceedings of the USENIX Workshop on Free and Open
  Communications on the Internet\/} (2013).

\bibitem{mediacom}
{\sc Bode, K.}
\newblock {Mediacom Injecting Their Ads Into Other Websites}.
\newblock \url{http://www.dslreports.com/shownews/112918}.

\bibitem{clayton2006ignoring}
{\sc Clayton, R., Murdoch, S.~J., and Watson, R.~N.}
\newblock Ignoring the great firewall of china.
\newblock In {\em Privacy Enhancing Technologies\/} (2006), Springer,
  pp.~20--35.

\bibitem{RFC5246}
{\sc Dierks, T., and Rescorla, E.}
\newblock The transport layer security ({TLS}) protocol version 1.2.
\newblock RFC 5246, August 2008.

\bibitem{duan2012hold}
{\sc Duan, H., Weaver, N., Zhao, Z., Hu, M., Liang, J., Jiang, J., Li, K., and
  Paxson, V.}
\newblock Hold-on: Protecting against on-path dns poisoning.
\newblock In {\em Proc. Workshop on Securing and Trusting Internet Names,
  SATIN\/} (2012).

\bibitem{RFC7230}
{\sc Fielding, R., and Reschke, J.}
\newblock Hypertext transfer protocol ({HTTP}/1.1): Message syntax and routing.
\newblock RFC 7230, June 2014.

\bibitem{herzberg2012security}
{\sc Herzberg, A., and Shulman, H.}
\newblock Security of patched dns.
\newblock In {\em Computer Security--ESORICS 2012}. Springer, 2012,
  pp.~271--288.

\bibitem{rtt}
{\sc Huffaker, B., Plummer, D., Moore, D., and claffy, k.}
\newblock {Topology discovery by active probing}.
\newblock In {\em Symposium on Applications and the Internet ({SAINT})\/} (Jan
  2002), pp.~90--96.

\bibitem{comcast}
{\sc Kearney, R.}
\newblock {Comcast caught hijacking web traffic}.
\newblock
  \url{http://blog.ryankearney.com/2013/01/comcast-caught-intercepting-and-altering-your-web-traffic/}.

\bibitem{kreibich2010netalyzr}
{\sc Kreibich, C., Weaver, N., Nechaev, B., and Paxson, V.}
\newblock Netalyzr: illuminating the edge network.
\newblock In {\em Proceedings of the 10th ACM SIGCOMM Conference on Internet
  Measurement\/} (2010), pp.~246--259.

\bibitem{leguay2005describing}
{\sc Leguay, J., Latapy, M., Friedman, T., and Salamatian, K.}
\newblock Describing and simulating internet routes.
\newblock In {\em NETWORKING 2005. Networking Technologies, Services, and
  Protocols; Performance of Computer and Communication Networks; Mobile and
  Wireless Communications Systems}. Springer, 2005, pp.~659--670.

\bibitem{levis2012collateral}
{\sc Levis, P.}
\newblock The collateral damage of internet censorship by dns injection.
\newblock {\em ACM SIGCOMM CCR 42}, 3 (2012).

\bibitem{mao2003towards}
{\sc Mao, Z.~M., Rexford, J., Wang, J., and Katz, R.~H.}
\newblock Towards an accurate {AS}-level traceroute tool.
\newblock In {\em Proceedings of the Conference on Applications, Technologies,
  Architectures, and Protocols for Computer Communications\/} (2003),
  pp.~365--378.

\bibitem{marczak2015analysis}
{\sc Marczak, B., Weaver, N., Dalek, J., Ensafi, R., Fifield, D., McKune, S.,
  Rey, A., Scott-Railton, J., Deibert, R., and Paxson, V.}
\newblock {An analysis of China's ``Great Cannon"}.
\newblock In {\em 5th USENIX Workshop on Free and Open Communications on the
  Internet (FOCI 15)\/} (2015).

\bibitem{mccanne1993bsd}
{\sc McCanne, S., and Jacobson, V.}
\newblock The {BSD} packet filter: A new architecture for user-level packet
  capture.
\newblock In {\em Proceedings of the Winter {USENIX} Conference\/} (1993),
  USENIX Association.

\bibitem{RIPEAtlas}
{\sc NCC, R.}
\newblock {RIPE Atlas}.
\newblock \url{https://atlas.ripe.net}.

\bibitem{RFC0793}
{\sc Postel, J.}
\newblock Transmission control protocol.
\newblock RFC 793, September 1981.

\bibitem{reis2008detecting}
{\sc Reis, C., Gribble, S.~D., Kohno, T., and Weaver, N.~C.}
\newblock Detecting in-flight page changes with web tripwires.
\newblock In {\em NSDI\/} (2008), vol.~8, pp.~31--44.

\bibitem{DefaultTTLs}
{\sc Siby, S.}
\newblock {Default TTL (Time To Live) Values of Different OS}.
\newblock \url{https://subinsb.com/default-device-ttl-values}, 2014.

\bibitem{Topolski2008}
{\sc Topolski, R.}
\newblock {NebuAd} and partner {ISPs}: Wiretapping, forgery and browser
  hijacking, June 2008.
\newblock \url{http://www.freepress.net/files/NebuAd_Report.pdf}.

\bibitem{RFC5925}
{\sc Touch, J., Mankin, A., and Bonica, R.}
\newblock The {TCP} authentication option.
\newblock RFC 5925, June 2010.

\bibitem{verkamp2012inferring}
{\sc Verkamp, J.-P., and Gupta, M.}
\newblock Inferring mechanics of web censorship around the world.
\newblock {\em Free and Open Communications on the Internet, Bellevue, WA,
  USA\/} (2012).

\bibitem{weaver2014here}
{\sc Weaver, N., Kreibich, C., Dam, M., and Paxson, V.}
\newblock Here be web proxies.
\newblock In {\em Passive and Active Measurement\/} (2014), Springer,
  pp.~183--192.

\bibitem{weaver2009detecting}
{\sc Weaver, N., Sommer, R., and Paxson, V.}
\newblock Detecting forged {TCP} reset packets.
\newblock In {\em NDSS\/} (2009).

\bibitem{rogers}
{\sc Weinstein, L.}
\newblock {Google Hijacked -- Major ISP to Intercept and Modify Web Pages}.
\newblock \url{http://lauren.vortex.com/archive/000337.html}.

\bibitem{xu2011internet}
{\sc Xu, X., Mao, Z.~M., and Halderman, J.~A.}
\newblock Internet censorship in china: Where does the filtering occur?
\newblock In {\em Passive and Active Measurement\/} (2011), Springer,
  pp.~133--142.

\bibitem{zhang2011inflight}
{\sc Zhang, C., Huang, C., Ross, K.~W., Maltz, D.~A., and Li, J.}
\newblock Inflight modifications of content: Who are the culprits.
\newblock In {\em Workshop of Large-Scale Exploits and Emerging Threats
  (LEET’11)\/} (2011).

\bibitem{Zimmerman2015}
{\sc Zimmerman, P.~T.}
\newblock Measuring privacy, security, and censorship through the utilization
  of online advertising exchanges.
\newblock Tech. rep., Princeton University, June 2015.

\end{thebibliography}

\appendix
\section{``Ack storm" due to TCP Injection} \label{ackstorm}
An ``Ack storm" occurs when the injected segment causes the receiver to send an acknowledgment for data bytes having sequence numbers  that were not yet sent by the peer. This acknowledgment is dropped by the peer, triggering it to respond by resending an earlier Ack, which may in turn trigger a retransmission by the receiver. The retransmitted segment will include again an acknowledgment for the yet to be sent sequence numbers and so forth. Such a ``ping-pong" exchange, if run long enough, will  cause the connection to timeout and reset. In many cases this is undesirable for the injector as it will interfere with the flow of traffic on the connection. An ``Ack storm" can subside if the peer eventually sends data bytes having sequence numbers that correspond to those of the forged data bytes injected by the third party. 

\section{Injection Detection Algorithm} \label{sec:injectdetectalg}
Algorithm~\ref{alg:racedetection} details the procedure for detecting packet races. This algorithm is executed by each worker process upon the receipt of a new packet. In the following, $CP$ denotes the currently received packet and $S$ denotes the set of packets received so far as part of the session of $CP$. $P(f)$ denotes the value of parameter $f$ of packet $P$. If parameter $f$ is a field of TCP or IP, it is denoted by the protocol and field names, e.g., $P(IP\_total\_length)$ denotes the value of the field Total Length in the IP header of packet $P$. The algorithm returns True if and only if a race is detected. 

\begin{algorithm*}
 \KwIn{CP, S}
 \ForEach{OP in S}{ \label{alg:line:foreach}
  \If{CP(t) - OP(t) $>$ MaxIntervalTime}{  \label{alg:line:MaxIntervalTime}
		continue\;
	}
	CP(headers\_size) = CP(IP\_header\_length) + CP(TCP\_data\_offset)*4\; \label{alg:line:1}
	OP(headers\_size) = OP(IP\_header\_length) + OP(TCP\_data\_offset)*4\; \label{alg:line:2}
	CP(payload\_size) = CP(IP\_total\_length) - CP(headers\_size)\;        \label{alg:line:3}
	OP(payload\_size) = OP(IP\_total\_length) - OP(headers\_size)\;        \label{alg:line:4}
	CP(top\_sequence\_number) = CP(TCP\_sequence\_number) + CP(payload\_size)\;  \label{alg:line:5}
	OP(top\_sequence\_number) = OP(TCP\_sequence\_number) + OP(payload\_size)\;  \label{alg:line:6}
	\If{CP(top\_sequence\_number) $>$ OP(TCP\_sequence\_number)}{                \label{alg:line:7}
		\If{OP(top\_sequence\_number) $>$ CP(TCP\_sequence\_number)}{              \label{alg:line:8}
			bottom\_overlap = MAX(CP(TCP\_sequence\_number), OP(TCP\_sequence\_number))\; \label{alg:line:9}
			top\_overlap = MIN(CP(top\_sequence\_number), OP(top\_sequence\_number))\;    \label{alg:line:10}
			\If{CP(TCP\_payload)[bottom\_overlap:top\_overlap] != OP(TCP\_payload)[bottom\_overlap:top\_overlap]}{  \label{alg:line:11}
				\KwRet{True}\;
			}
		}
	}
 }
  \KwRet{False}\;
 \caption{Race detection algorithm} \label{alg:racedetection}
\end{algorithm*} 

In Algorithm~\ref{alg:racedetection}, line~\ref{alg:line:foreach} iterates over the previously received packets of the current session. Line~\ref{alg:line:MaxIntervalTime} verifies that the two considered packets have been received within a time interval that does not exceed the parameter $MaxIntervalTime$. Lines~\ref{alg:line:1} and \ref{alg:line:2} compute the total lengths of the TCP and IP headers of each of the two packets. Lines~\ref{alg:line:3} and \ref{alg:line:4} compute the payload size of each of the two packets. Lines~\ref{alg:line:5} and \ref{alg:line:6} compute the TCP sequence number of the last byte delivered in the payload  in each of the two packets. Lines~\ref{alg:line:7} and \ref{alg:line:8} check for a sequence number overlap between the two packets. Line~\ref{alg:line:11} checks whether the overlapped payload is different. If it is, a race is detected and the algorithm returns True. 

To avoid false positives, we did not consider the following packets (not shown in Algorithm~\ref{alg:racedetection}):
\begin{enumerate}
	\item Checksum errors -- packets that have checksum errors either in the TCP or IP headers will clearly have a different payload than that of their retransmission.
	\item TCP reset -- reset packets can carry data payloads for diagnostic messages which are not part of the regular session's byte stream.
\end{enumerate}

\section{False Positives} \label{sec:falsepositives}
There were numerous events in which the algorithm in Appendix~\ref{sec:injectdetectalg} identified a race that was not due to a forged packet injection into the TCP session. In the following we describe these events and why they occur:

\paragraph{Retransmissions with different content} As per the TCP specification \cite{RFC0793}, the payload of retransmitted segments must have the same content as the payload of the original segment. In practice, however, this is not always the case, and retransmitted segments sometimes carry slightly different content, for the following reasons:
	\begin{itemize}
		\item Load balancing -- some websites serve HTTP requests using more than one server. Usually, a front-end load balancer redirects the HTTP requests according to the current load on each web server. It is sometimes desirable that the same server serve all HTTP requests coming from the same client. To facilitate this, the first HTTP response sent to a client sets a cookie containing the identity of the server chosen to serve the client from now on. Subsequent requests from that client will include this server ID and allow the load balancer to redirect those requests to that server. If the first HTTP response needs to be retransmitted, some load balancers might, at the time of the retransmission, choose a different web server than the one they originally chose when the response was first transmitted. This results in a different cookie value set in the retransmitted response. Examples of websites that exhibit such behavior are \url{wiley.com} and \url{rottentomatoes.com}.

		\item Accept-Ranges HTTP header -- the HTTP 1.1 specification \cite{RFC7230} allows a client to request a portion of a resource by using the Range header in the HTTP request. It may do so in cases where the web server has indicated in previous responses its support of such range requests. Such support is indicated by the Accept-Ranges header. We observed cases where a web server sent an HTTP response which included  'Accept-Ranges:  none', indicating that the server is unwilling to accept range requests, while in a retransmission of the same response the header was replaced by 'Accept-Ranges: bytes', indicating that it is willing to accept range requests having units of bytes. This happened when the retrieved resource spanned multiple TCP segments. Presumably, the intention of the server is to allow the client to retrieve a portion of a resource when network loss is high. Examples of websites that exhibit such behavior are \url{sagemath.org} and \url{nih.gov}. Furthermore, such behavior was exhibited by several types of web servers, including Apache, nginx and IIS.
		
		\item Non-standard HTTP headers -- we have observed that in some web applications that use non-standard HTTP headers (namely, headers that begin with \mbox{'x-'}), a retransmission of an HTTP response has different values for these headers than their value in the initial response. For example, Amazon's S3 service includes in every response the headers 'x-amz-id-2' and 'x-amz-request-id', which help to troubleshoot problems. These headers have a unique value for each response even if it is a retransmission.
		
	\end{itemize}
	\paragraph{Retransmissions with different sequence numbers} For a few websites we encountered sessions in which a retransmitted TCP segment started with a sequence number that was offset by 1 compared to the sequence number of the original segment. This might occur due to a bug that caused the unnecessary incrementation when a FIN segment was sent between the original and retransmitted segment. There were no indications in the HTTP responses as to the type of software executed by those web servers. This unnecessary incrementation might also be an artifact of a middle-box that serves the traffic to those servers.  An example of a website that exhibits such behavior is \url{www.knesset.gov.il}.
	\paragraph{Non-compliant TCP traffic} We encountered many TCP sessions (over port 80) which do not appear to have originated from TCP-compliant nodes. There was no proper 3-way handshake to open the session, the acknowledgment did not correspond to the actual received bytes, flags were set arbitrarily, and the sequence numbers were not incremented consecutively. This last point led our monitoring system to flag many of these sessions as injected sessions. Many of these sessions included only unidirectional incoming traffic that originated from a handful of networks primarily residing in hosting providers (such as GoDaddy and Amazon). We suspect that these are communication attempts by a command and control server to its bots. However, we have no proof of this.

\begin{algorithm*}
 \KwIn{CP, S}
	\If{Check\_Race(CP,S(Suspicious\_Queue))}{ 
		Block suspicious packet\;
	} 
	Suspicious = False\;
	\If{abs(CP(IP\_TTL)-S(Average\_TTL)) $>$ 1} {
		Suspicious = True\;
	}
	Lower\_ID\_Boundary = (S(Last\_ID) - 10)\%$2^{16}$\;
	Upper\_ID\_Boundary = (S(Last\_ID) + 5000)\%$2^{16}$\;
	\If{CP(IP\_ID) $<$ Lower\_ID\_Boundary or CP(IP\_ID) $>$ Upper\_ID\_Boundary} {
		Suspicious = True\;
	}
	\If{Suspicious == True} {
		S(Suspicious\_Queue).append(CP)\; 
	}
	\Else {
		Update S(Average\_TTL) with CP(IP\_TTL)\;
		S(Last\_ID) = CP(IP\_ID)\;
		Accept CP\;
	}
 \caption{Mitigation algorithm} \label{alg:mitigation}
\end{algorithm*} 

\section{Attempts to Mimic the Identification Values of the Legitimate Packet} \label{sec:injectedid}
In the following we account for some of the failed attempts we observed in which the injecting entity tried to mimic the identification value of the legitimate packet. Note that in order to increase the chances of winning the race with the legitimate packet, the forged packet is injected well before the injecting entity has a chance to inspect it. For this reason the injecting entity can not simply copy the identification value of the legitimate packet to the forged one.

\begin{enumerate}
	\item Duplicate ID with a packet from the server -- in some cases the injecting entity tries to mimic the identification values of the packets sent by the server to make the forged packet less conspicuous. Sometimes this is done rather carelessly by simply copying the identification number of one of the packets the server already sent (not the legitimate packet the entity wishes to forge). This means that the client receives two IP packets from the server having the exact same identification number. This situation is highly unlikely to occur without the intervention of a third party in the session, as the IP layer of the server must make sure that each packet in the session has a unique identification value. 
	\item Duplicate ID with a packet from the client -- in different attempts to, perhaps, mimic the identification values of the packets sent by the server, some injectors simply copy an identification value from the HTTP request packet that triggered the response. Since this packet is, of course, sent by the client, the injector cannot achieve its goal; the identification values of the packets sent by the client are completely independent of those sent by the server. We can use this to our advantage. It is possible but unlikely that two packets -- one sent by the server and the other by the client -- have the same identification value. 
	\item Swapped bytes of an ID in packets coming from the client -- we noticed that at least one injector that aims to copy the identification value from a packet coming from the client (as described in the previous rule), does so in such a way that the two bytes of the copied values are swapped. For example, if the identification value of a packet coming from the client is 0xABCD, then the identification value of the injected packet will be 0xCDAB. This is probably due to a bug of the injector\footnote{Most likely the bug is a case of big endian/little endian confusion.}. Occurrence of such an event is highly unlikely without third-party intervention. 
\end{enumerate}

\section{Improved Mitigation Algorithm} \label{sec:mitigationalg}

Algorithm~\ref{alg:mitigation} details the proposed mitigation algorithm. The algorithm is executed upon the receipt of a new incoming packet -- $CP$. As in Algorithm~\ref{alg:racedetection} above, $S$ denotes the session of $CP$. $P(f)$ denotes the value of parameter $f$ of packet $P$. If parameter $f$ is a field of TCP or IP, it is denoted by the protocol and field names, e.g., $P(IP\_ID)$ denotes the value of the field Identification in the IP header of packet $P$. 

The algorithm maintains a queue of packets that are suspected of being forged. The incoming packet is first checked against the suspicious packets for a race. If a race is detected, the suspicious packet is blocked. Afterward, the TTL of the incoming packet is compared against the average of TTL values of the previous packets received in the same session. If the difference is larger than 1, then the packet is marked as suspicious. The packet is also marked as suspicious if its Identification value is higher than 5000 plus the Identification value of the previously received packet of the session or lower than that value minus 10. The rationale behind this comparison is that we generally expect the Identification values of the session be monotonically increasing, except in cases of packet reordering. If the packet is marked as suspicious it is enqueued to the suspicious queue for 200ms. If the packet is not suspicious the value of the average TTL and last ID are updated and the packet is accepted. 

Note that a race will not be identified if the injected packet arrives \emph{after} the legitimate one. This is because the legitimate packet will not be delayed, and once the inject packet is received it will not be checked for a race against the legitimate one. Nonetheless, this does not compromise the security of the client since in this case the content of the injected packet will not be accepted by the client's TCP layer.

\end{document}